\documentclass{article}

\PassOptionsToPackage{numbers, compress}{natbib}
\usepackage[preprint]{neurips_2026}

\usepackage[utf8]{inputenc} 
\usepackage[T1]{fontenc}    
\usepackage{hyperref}       
\usepackage{url}            
\usepackage{booktabs}       
\usepackage{amsfonts}       
\usepackage{nicefrac}       
\usepackage{microtype}      
\usepackage{xcolor}         
\usepackage{graphicx}
\usepackage{caption}
\usepackage{amsmath}
\usepackage{algorithm}
\usepackage{algorithmic}

\title{\textsc{AgentWeave}: Efficient Distributed Agent Serving via Flow Decomposition}

\author{
\textbf{Kaibin Guo}$^{1}$,
\textbf{Pengtu Li}$^{1}$,
\textbf{Zicong Hong}$^{2}$\thanks{Corresponding author.},
\textbf{Zhiyuan Fang}$^{1}$,\\
\textbf{Wuhui Chen}$^{1}$,
\textbf{Rachid Guerraoui}$^{2}$,
\textbf{Anne-Marie Kermarrec}$^{2}$\\
\normalfont
$^{1}$Sun Yat-sen University,
$^{2}$EPFL\\
\texttt{\{guokb,lipt5,fangzhy27\}@mail2.sysu.edu.cn},\texttt{chenwuh@mail.sysu.edu.cn},\\
\texttt{\{zicong.hong,rachid.guerraoui,anne-marie.kermarrec\}@epfl.ch}
}

\begin{document}

\maketitle

\begin{abstract}
Agent applications are evolving into workflows composed of large language model (LLM) calls and tool invocations, termed \emph{modules}. Existing distributed agent systems typically adopt a disaggregated architecture, deploying different modules on separate devices.
While this enables pipeline parallelism, imbalanced module latencies introduce severe pipeline bubbles that limit throughput.
A natural alternative is an aggregated architecture that co-deploys all modules across all devices, eliminating bubbles by keeping every device continuously active.
However, this introduces \emph{request blocking}: short requests are forced to progress in lockstep with longer ones in the same batch, and newly arriving requests must wait until the entire ongoing batch completes the full workflow.
To address this, we propose \textsc{AgentWeave}, an efficient distributed serving system for agent applications built on the aggregated architecture.
Its core mechanism, \emph{flow decomposition}, decomposes long agent requests into fine-grained execution units while preserving agent semantics, allowing short and newly arriving requests to proceed without being stalled by long-running ones.
Complementing this, a \emph{flow-adaptive scheduler} dynamically balances the throughput gains of decomposition against its scheduling overhead.
Experiments across four representative agent applications demonstrate that \textsc{AgentWeave} improves throughput by $1.81\times$, reduces average latency by $2.45\times$, and reduces P90 latency by $2.04\times$ over state-of-the-art baselines.
\end{abstract}




\section{Introduction}
\label{intro}

Large language model (LLM)-based agents~\cite{agent} have emerged as a compelling paradigm for tackling complex real-world tasks. 
Yet building capable agents is non-trivial, as a single LLM~\cite{gpt,bert} alone is often insufficient: it may lack up-to-date knowledge~\cite{rag_2023,webgpt}, struggle to reason over long contexts~\cite{lost,longbench,found}, and cannot reliably interact with external data sources~\cite{toolformer,apibank}.
To overcome these limitations, modern agents augment LLMs with specialized tools that enable them to navigate external environments and solve complex tasks more effectively~\cite{toolformer, react, tool2}.
We adopt \emph{module} as a unified abstraction encompassing both LLM calls and tool invocations; under this abstraction, a modern agent is naturally organized as a workflow of modules.

\begin{figure}[t]
  \centering
  \includegraphics[width=0.75\columnwidth]{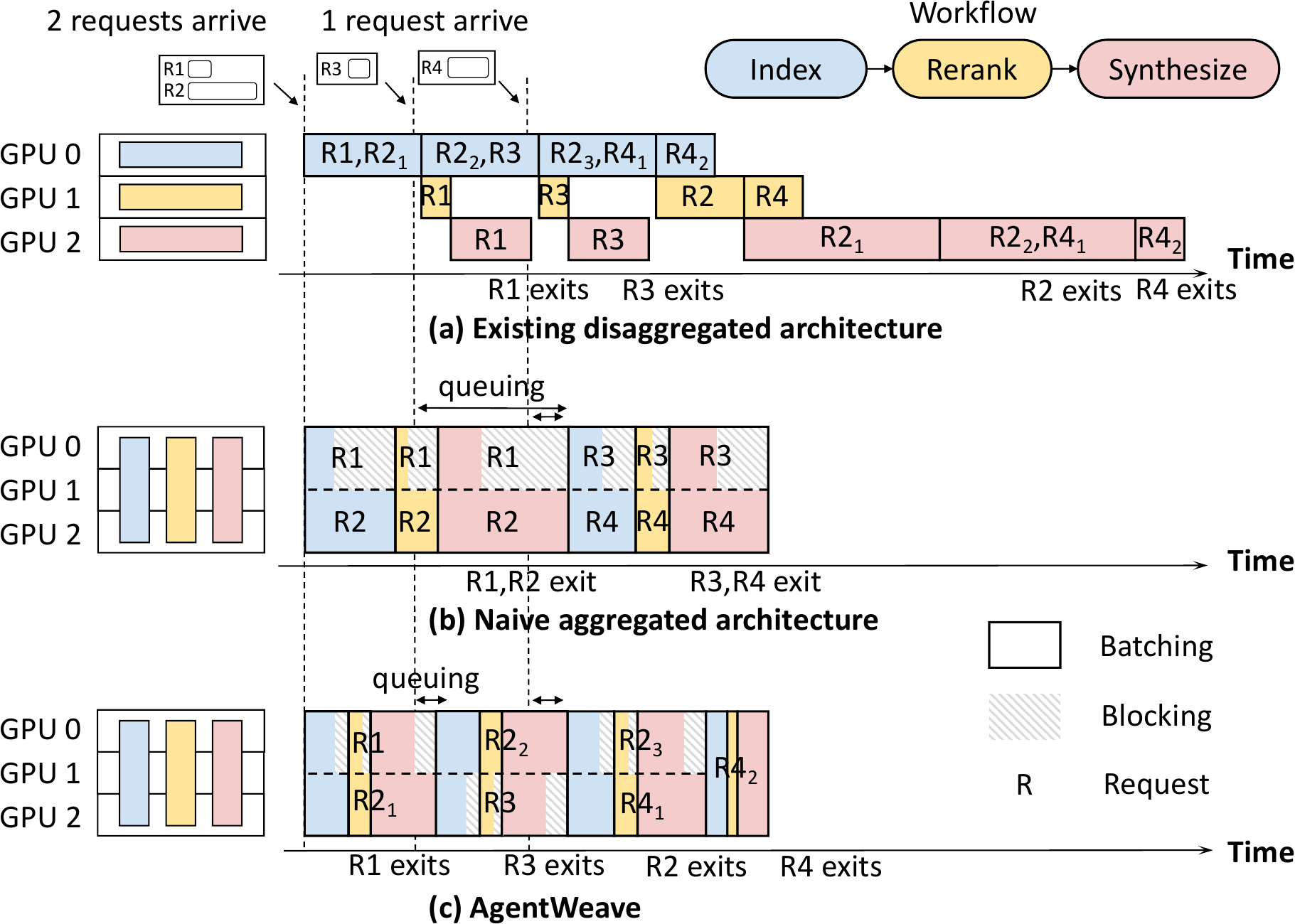}
  \caption{Comparison of execution paradigms for agent serving on simplified RAG, drawn to realistic scale, consisting of the indexing, reranking, and synthesizing modules that dominate RAG latency.}
  \label{fig:compare}
  \vspace{-5pt}
\end{figure}

Existing distributed serving systems~\cite{ayo, autellix, parrot, vllm, sglang, continuum} adopt a \emph{disaggregated} architecture for agent serving, where different modules are deployed on separate devices and managed independently. 
This design naturally forms a pipeline that allows modules to overlap execution across requests. 
However, it fundamentally suffers from \emph{pipeline bubbles}: because module latencies are highly imbalanced in practice, faster modules are forced to idle while waiting for slower ones, leaving substantial dead time in the pipeline.
Figure~\ref{fig:compare}(a) illustrates this architecture, where three modules are deployed on \(GPU0\), \(GPU1\), and \(GPU2\), respectively. 
When short requests \(R1\), \(R3\) and long requests \(R2\), \(R4\) arrive, modules execute in a pipelined fashion, yet the resulting pipeline bubbles severely degrade overall serving performance.

A natural alternative is the \emph{aggregated} architecture, which co-deploys all modules across all devices, fully utilizing every device at each step of the workflow. 
This eliminates pipeline-induced bubbles and accelerates per-request execution. 
As illustrated in Figure~\ref{fig:compare}(b), all three modules are deployed across all devices.
\(R1\) and \(R2\) are batched and processed through three modules, exiting together after completing the entire agent workflow. \(R3\) and \(R4\) also follow the same pattern. Compared with Figure~\ref{fig:compare}(a), requests complete earlier under this design, yielding lower latency and higher throughput.


Despite these advantages, the aggregated architecture introduces a critical inefficiency: \emph{request blocking}.
As shown in Figure~\ref{fig:compare}(b), although short request \(R1\) requires far less computation in the indexing stage, it cannot advance to reranking earlier, because all devices remain committed to indexing for the entire batch until long request \(R2\) finishes. 
Consequently, short requests are forced to exit in lockstep with long ones, while newly arriving requests such as \(R3\) must wait until the ongoing batch (\(R1\), \(R2\)) completes the full workflow before they can begin execution. 

We therefore present \textsc{AgentWeave}, an efficient distributed serving system for agent applications. 
Our key idea is to decompose long requests at the agent level. As shown in Figure~\ref{fig:compare}(c), we decompose \(R2\) into smaller execution units \(R2_1\), \(R2_2\), and \(R2_3\). So \(R1\) can exit after executing with \(R2_1\), rather than being coupled with the entire \(R2\). New requests such as \(R3\) can also enter between two units of a long request and complete earlier. This reduces request blocking while retaining the benefits of the aggregated architecture. To practically realize this idea, we further address two technical challenges.

\begin{figure}[t]
  \centering
  \includegraphics[width=\columnwidth]{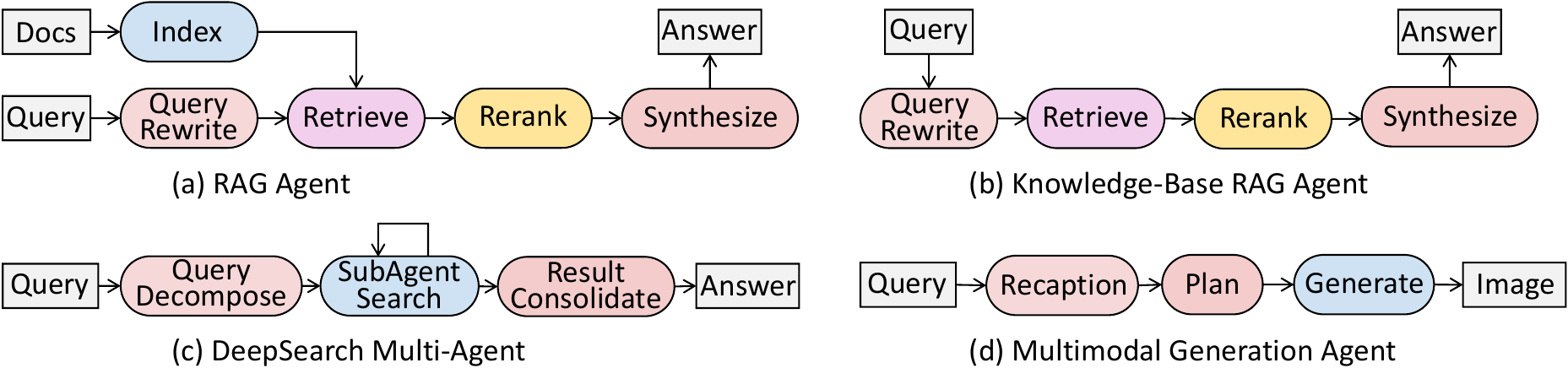}
  \caption{Representative agent applications, showing that modern agents are typically organized as a workflow of modules with inter-module execution dependencies.}
  \label{fig:workflow}
\end{figure}

\textbf{Challenge 1: How to decompose a request without breaking agent semantics.} 
This requires preserving two semantics constraints: execution workflow, i.e., the module execution structure shown in Figure~\ref{fig:workflow}; and complete-input requirement, where a module requires a complete input to produce correct output, e.g., reranking requires the full retrieved documents to select the global top-$k$ results. 

\textsc{AgentWeave} introduces \emph{flow decomposition} to decompose each request while preserving agent semantics. 
It develops an abstraction called \emph{flows} (i.e., module's input or output) as the decomposable unit, and rewrites the module dependency into a flow dependency graph. It decomposes flows along valid flow dependency, and inserts supplement dependencies to satisfy complete-input requirement. 

\textbf{Challenge 2: How to control decomposition cost.}
Decomposition reduces blocking, but also introduces scheduling and flow-management costs. Over-decomposition can accumulate cost and significantly increase the latency of long requests.

\textsc{AgentWeave} introduces a \emph{flow-adaptive scheduler} to control this cost. Instead of using a fixed decomposition size, it determines the size online by balancing the
throughput gain against the cost penalty, and then constructs the corresponding batch for execution. This guarantees the decomposition gains without
increasing long-tail latency.

We implement a prototype of \textsc{AgentWeave} on top of SGLang\cite{sglang}, and evaluate it on four representative agent applications, across diverse model-hardware configurations. Evaluation shows that it improves throughput by 1.81$\times$, reduces average and P90 latency by 2.45$\times$ and 2.04$\times$, respectively.

\section{Background and Motivation}

\subsection{Agent Applications}
\label{motivation:workflow}

Agents have become increasingly popular in real-world applications.
Figure~\ref{fig:workflow} shows four representative examples (see Appendix~\ref{sec:agent_primer} for details).
In a retrieval-augmented generation (RAG) agent~\cite{ayo,rag_2020,rag_2023}, an LLM rewrites the query, invokes an embedding tool to index documents and retrieve relevant ones, a reranking tool to refine results, and finally synthesizes the answer, thereby enabling better knowledge-grounded generation.
A knowledge-based (KB) RAG agent~\cite{rag_2020,hedrarag} follows the RAG agent but retrieves candidates from a pre-built knowledge base instead of indexing new documents.
In a DeepSearch agent~\cite{deepsearch,manusearch}, an LLM decomposes a query into subplans, coordinates search subagents tools to collect and summarize evidence, and consolidates the results into a structured response.
In a multimodal generation agent~\cite{rpg,llmgrounded}, an LLM refines the query, plans the scene layout, and invokes a diffusion generator to produce the final image.

\subsection{Inefficiency of Disaggregated Architecture}
\label{disaggregated_inefficiency}


Existing distributed serving systems~\cite{autellix,ayo,vllm,sglang} adopt a disaggregated architecture in Figure~\ref{fig:compare}(a) for agent applications, and primarily pursue two optimization directions: workflow orchestration and module serving.
The first direction~\cite{ayo} optimizes workflow by splitting modules into finer-grained units and overlapping their execution to shorten the latency of a request.
The second direction~\cite{autellix,parrot,vllm,sglang} optimizes individual modules serving, most commonly the LLM-based modules, by reducing queueing delay or eliminating redundant computation and memory.

\begin{figure}[ht]
  \centering

  \begin{minipage}[t]{0.49\columnwidth}
    \vspace{0pt}
    \centering
    \includegraphics[width=\linewidth]{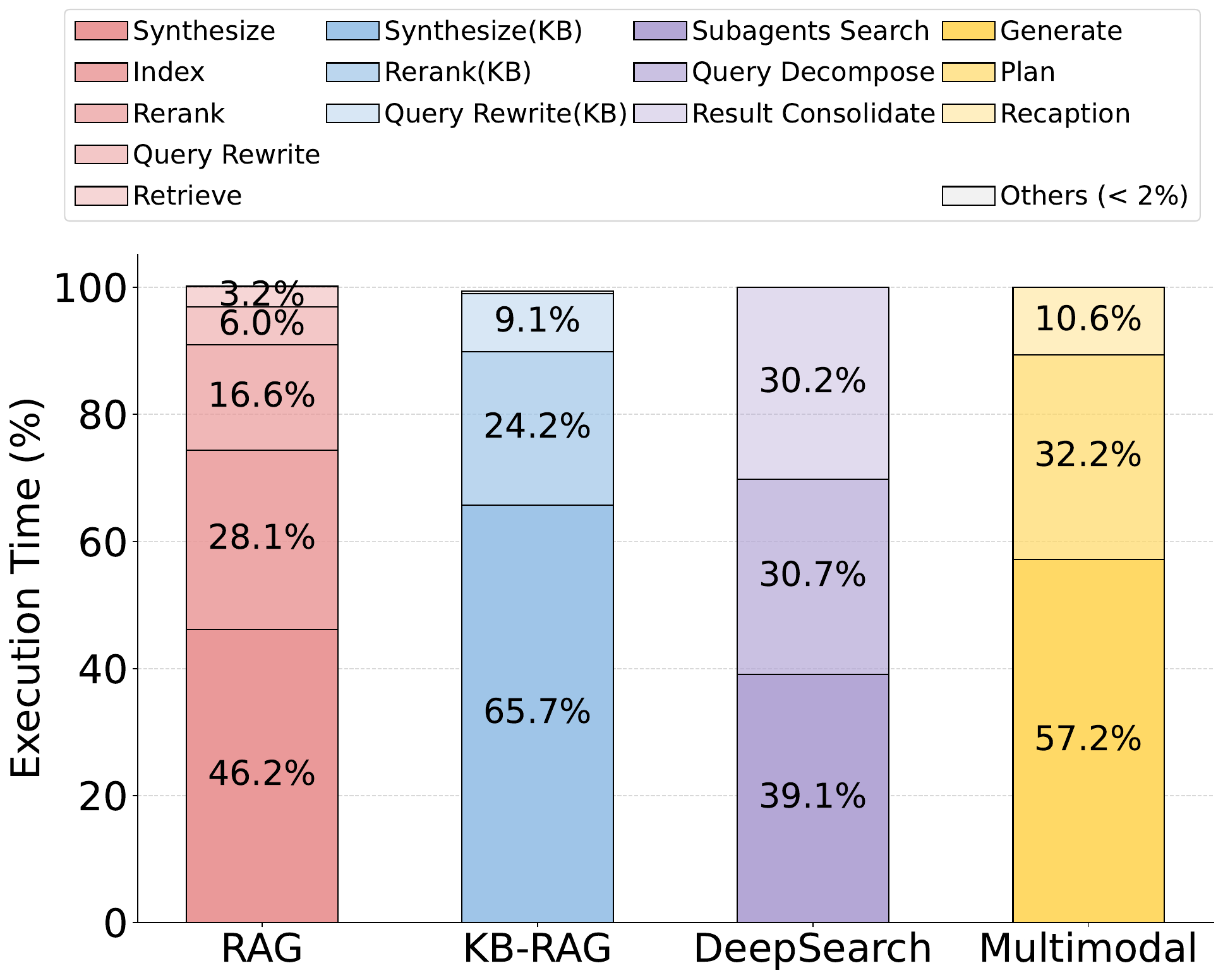}
    \captionof{figure}{Execution time breakdown of representative agent applications, including RAG, KB-RAG, DeepSearch and Multimodal. The results illustrate module latency imbalance.}
    \label{fig:percentage_breakdown}
  \end{minipage}%
  \hfill
  \begin{minipage}[t]{0.47\columnwidth}
    \vspace{0pt}
    \centering
    \includegraphics[width=\linewidth]{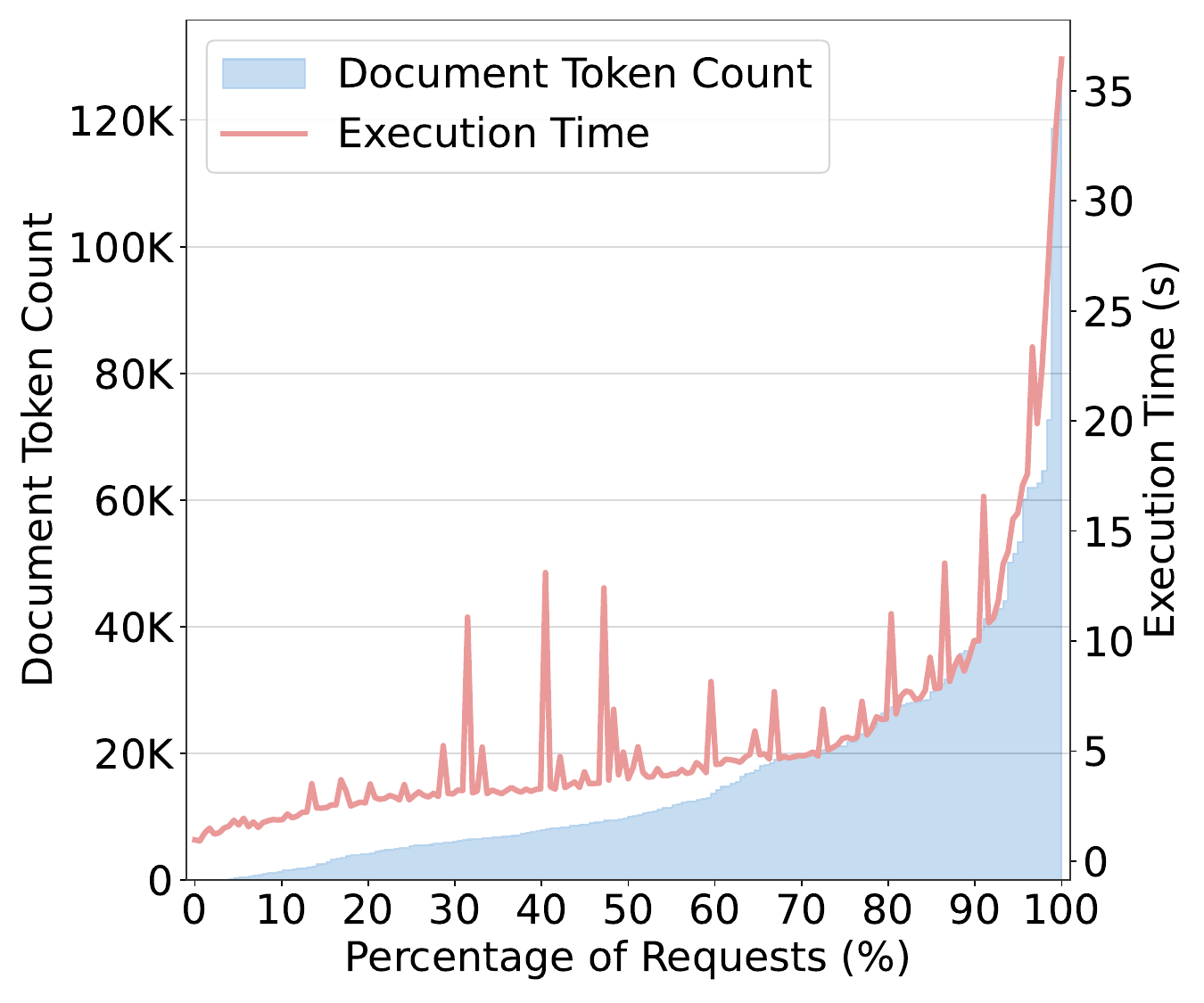}
    \captionof{figure}{Execution-time heterogeneity of agent requests in RAG agent. Different requests carry different document, which lead to substantially different end-to-end execution times.}
    \label{fig:hol}
  \end{minipage}
\end{figure}

However, both directions still suffer from pipeline bubbles~\cite{gpipe,pipedream} caused by imbalanced module latencies. Figure~\ref{fig:percentage_breakdown} illustrates this imbalance: execution time is distributed unevenly across modules in all four agent applications, with latency skew reaching 41.5\% in KB-RAG, where LLM-based synthesis accounts for 65.7\% and reranking by the rerank model accounts for only 24.2\%. This imbalance leaves severe bubble time in the pipeline, reducing effective overlap and limiting throughput. 



\subsection{Aggregated Architecture: Benefits and Inefficiency}
This motivates an aggregated architecture in Figure~\ref{fig:compare}(b), which co-deploys all modules of an agent across all devices. It removes pipeline and fully utilizes devices to process each module, therefore avoiding bubbles and additionally accelerating the request execution. These benefits can reduce latency and improve system throughput.

However, this architecture still suffers from request blocking. 
Real-world requests can vary significantly in execution time, as users or platforms choose different inputs or hyperparameters to satisfy budget, accuracy, and SLO requirements~\cite{deepsearch,azure,cohere,google}. Figure~\ref{fig:hol} shows this variation for RAG agent on TriviaQA~\cite{triviaqa}, where execution time significantly differs across request percentiles, with the gap even exceeding 20$\times$. Short and long requests can be batched and advance together, causing request blocking, and substantially limit the latency and throughput gains of the aggregated architecture.

\subsection{Decomposition and Challenge}
\label{motivation:flow}
The blocking stems from executing a long request as one coarse-grained unit. A naive solution is to decompose it into smaller execution units. 
In this way, short requests can be executed with a small execution unit and exit sooner, rather than being coupled with the entire long ones.  
New requests can enter between two consecutive units of the long request, rather than waiting for its entire completion.


We observe that request execution mainly depends on the input or output data flowing through modules. 
This is evident in Figure~\ref{fig:hol}: as the document directly determines the indexing cost and indirectly affects downstream execution by propagating through retrieval, reranking, and synthesis, it largely determines the end-to-end execution time.
Thus, a long RAG request can be decomposed by splitting its documents to trigger only part of the original execution. 
By decomposing such data, we can decompose the corresponding executions and obtain smaller execution units.


However, decomposition presents two challenges. 
\textbf{First, preserving agent semantics}.
Semantics impose strict execution workflow, as modules must follow the specified order;
and complete-input requirement, as some modules require complete input to produce correct output. 
Naive decomposition may violate these constraints, causing downstream modules to execute out of order or consume partial input when complete input is required. 
Thus, decomposition must preserve agent semantics to avoid degrading the answer quality.
\textbf{Second, controlling decomposition cost}. 
Although decomposition reduces blocking, it introduces scheduling and management cost. 
For long requests, these costs may accumulate across scheduling rounds, significantly increasing latency. 
Thus, decomposition must control its granularity to avoid long-tail latency.

\section{Method}
We present \textsc{AgentWeave}'s overall architecture (§\ref{overview}) and flow abstraction (§\ref{flow_abstraction}), followed by its two key techniques: (1) flow decomposition (§\ref{decomposition}) to provide a semantic-preserving decomposition method, and (2) a flow-adaptive scheduler (§\ref{scheduler}) to guarantee decomposition gains.

\begin{figure}[ht]
  \centering
  \includegraphics[width=0.83\textwidth]{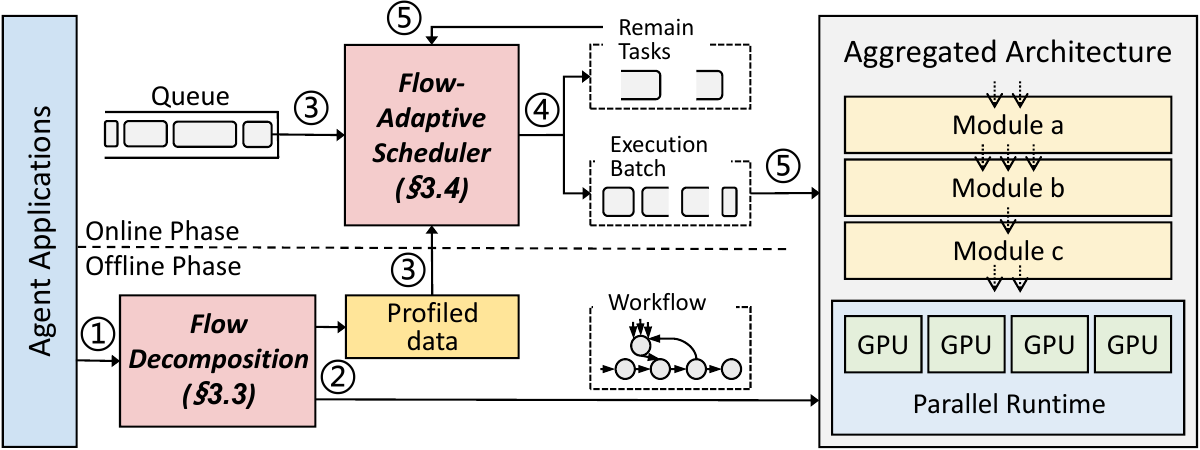} 
  \caption{System Overview of \textsc{AgentWeave}.}
  \label{fig:overview}
\end{figure}

\begin{figure}[t]
  \centering
  \includegraphics[width=\textwidth]{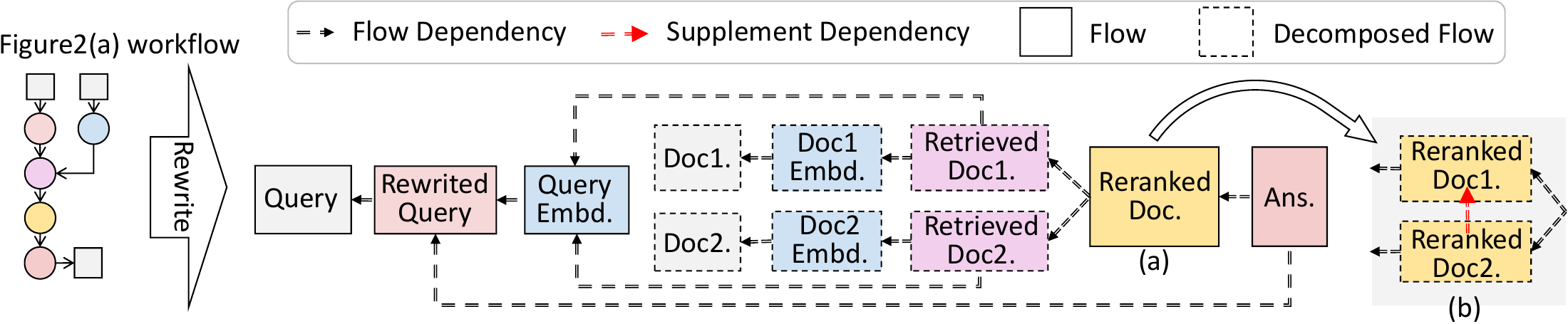} 
  \caption{Flow decomposition on RAG agent in Figure~\ref{fig:workflow}(a). We illustrate other agents in Appendix\ref{other_decomposition}.}
  \label{fig:decomposition}
\end{figure}

\subsection{Overview}
\label{overview}
Figure~\ref{fig:overview} shows the overview of \textsc{AgentWeave}.
In the offline phase, the system registers an agent application (\textcircled{1}), analyzes its workflow to derive a semantics-preserving decomposition method (\textcircled{2}), and profiles module execution time and decomposition costs for online scheduling. In the online phase, the scheduler dequeues requests and determines the decomposition size based on the offline profiling results (\textcircled{3}). It then constructs the execution batch (\textcircled{4}), dispatching the batch to the aggregated engine and returning remaining tasks to the scheduler for later rounds (\textcircled{5}).

\subsection{Flow Abstraction}
\label{flow_abstraction}

\textbf{Flow as the Unit of Decomposition.} 
A module's input or output data determines its required computation.
We term such data a \emph{flow}, and decompose a long request by decomposing flows into subflows for module execution.
We further identify which flow can be decomposed.



A flow can be decomposed only when (1) it can be split into independent subflows, and (2) each subflow is a correct subset of the flow produced by its module on the complete input.
Otherwise, the flow should be processed as a whole.
In Figure~\ref{fig:decomposition}, for example, document embeddings can be decomposed because each embeddings subflow is independent and represents a correct subset of the full embeddings produced by indexing on complete documents.
In contrast, the query is not decomposable because partial queries are dependent on each other to express the original user intent.


Some flows satisfy the first condition but not the second.
They can be decomposed into independent subflows, but each subflow may not be a correct subset.
For reranking (a) example in Figure~\ref{fig:decomposition}, reranked documents can be split into independent document subflows, but it requires the complete retrieved documents as input to select the global top-$k$ results, making subflow not a correct subset result, thus not decomposable.
We treat such flows as conditionally decomposable and decompose them while preserving the complete-input requirement through supplement dependency in Section~\ref{decomposition}.





\textbf{Flow Dependency Graph.}
The flow dependency graph is a flow representation of the module workflow, where nodes denote data flows and edges denote how outputs are derived from inputs by modules. In this graph, each module's output flow depends on its input flows, thus, the original execution order is represented as flows dependency, as illustrated by the RAG example in Figure~\ref{fig:decomposition}. As flow is the appropriate unit of decomposition, this reformulation makes decomposition explicit.

\subsection{Flow Decomposition}
\label{decomposition}
Flow decomposition enables long requests to be split into smaller execution units while preserving agent semantics. The system constructs decomposition based on above abstraction.

\textbf{Decomposition Construction.}
The system first rewrites the workflow into a flow dependency graph, then traces and splits the decomposable flows along the dependency from the final answer. In Figure~\ref{fig:decomposition}, it decomposes the retrieved documents into subflows, and further propagates to document embeddings and raw documents. It keeps other flows intact, as they are not decomposable. This step determines which flows are decomposed and how they propagate along the valid dependency.


\textbf{Supplement Dependency Construction.}
The system then splits conditionally decomposable flows and inserts supplement dependencies to satisfy complete-input requirement. A supplement dependency connects a later subflow to an earlier one, allowing the later subflow to receive the accumulated result from previous subflows and continue the original complete-input computation. For reranking (b) in Figure~\ref{fig:decomposition}, the system inserts a supplement dependency from reranked document subflow $2$ to subflow $1$.
The second reranking then consumes both the retrieved documents subflow $2$ and the accumulated reranked result from subflow $1$, producing the correct top-\(k\) result.
This step splits conditionally decomposable flows while preserving their complete-input requirement.


\textbf{Decomposition of a Request.}
At runtime, the system decomposes a long request by splitting the most downstream decomposed flow in the graph, i.e., documents in RAG. 
Each subflow is consumed to produce output subflows along the reverse direction of flow and supplement dependencies. 
Since each module consumes or produces only a subflow rather than the complete flow, its execution is shorter, and the execution triggered by one subflow forms a smaller execution unit for scheduling.

This runtime decomposition is controlled by the decomposition size, which sets the subflow size for each request in a scheduling round.
A larger decomposition size increases per-round execution time, while a smaller one creates more execution units and introduces more overhead.
Thus, the system profiles execution time and overhead under diverse decomposition sizes for online size selection.

\subsection{Flow-Adaptive Scheduler}
\label{scheduler}
The flow-adaptive scheduler determines the decomposition size online in each scheduling round based on the current queue state and profiled data. The goal is to maximize system throughput while controlling the long tail latency caused by decomposition overhead. 

\textbf{Problem Formulation.}
We formulate decomposition size selection as an online optimization problem. At each scheduling round, let $\mathcal{R} = \{r_1, r_2, \dots, r_N\}$ be the requests for current execution; $w_i$ be the remaining workload of request $r_i$; $S$ be a candidate decomposition size; $\mathcal{S}$ be the candidate set.

Given a decomposition size $S$, current round total time is
\(
T_{\text{total}}(S) = T_{\text{exec}}(S, \mathcal{R}) + O(S),
\)
where $T_{\text{exec}}(S, \mathcal{R})$ is the effective execution time under decomposition size $S$, and $O(S)$ is the decomposition overhead introduced in this round. Both terms are estimated from the offline workload profiles.

Throughput is measured by completed requests per unit time, but this discrete metric ignores partial progress made by unfinished requests. 
We therefore use a continuous throughput-gain objective:
\[
J_{\text{gain}}(S) = 
\frac{\sum_{r_i \in \mathcal{R}} \min\left(1, S / w_i\right)}
{T_{\text{total}}(S)},
\]
where each request contributes $1$ if it completes in the current round, and otherwise contributes $S/w_i$ to represent the fraction of its workload completed.

Decomposition incurs overhead each round, which accumulates over rounds and increases request completion time. Since the decomposition sizes of future rounds are unknown, we estimate this cost by applying the current size $S$ to the remaining workload and summing the resulting overhead:
\[
O_{\text{total}}(S) = \sum_{r_i \in \mathcal{R}} \left\lceil w_i/S \right\rceil \cdot O(S).
\]

Since long requests are decomposed across more rounds, accumulated overhead can increase tail latency. The scheduler therefore selects the decomposition size $S$ to maximize throughput gain while controlling accumulated decomposition overhead.



\textbf{Problem Solving.}
This scheduling problem corresponds to a multi-objective optimization problem~\cite{nonlinear}, and we follow the classical scalarization method~\cite{nonlinear}, as it provides trade-off-specific optimality and low decision overhead for online scheduling. We scalarize the throughput-gain and accumulated-overhead objectives into a single objective:
\[
S^{*}
=
\arg\max_{S \in \mathcal{S}}
\mathrm{Score}(S),
\quad
\mathrm{Score}(S)
=
J_{\mathrm{gain}}(S)
-
\lambda \cdot O_{\mathrm{total}}(S),
\quad
\lambda > 0.
\]



In classical scalarization, \(\lambda\) is a trade-off parameter that reflects the relative preference between objectives. 
We can obtain a fixed \(\lambda\) by offline system profiling, but a fixed preference cannot adapt to the online requests, where the relative importance of throughput gain and accumulated overhead changes with the queue state.
We therefore estimate \(\lambda\) online by quantifying the throughput loss caused by additional overhead as the current preference.

Let $\bar{T}_{\text{exec}}$ be the system average execution time. The ideal throughput is
\(
J \approx {|\mathcal{R}|} / {\bar{T}_{\text{exec}}}.
\)
The sensitivity of throughput to additional latency is
\(
{\partial J}/{\partial T} = -{|\mathcal{R}|}/{\bar{T}_{\text{exec}}^2}.
\)
We set the trade-off parameter as
\(
\lambda = {|\mathcal{R}|}/{\bar{T}_{\text{exec}}^2},
\)
which converts the accumulated cost into an equivalent throughput loss, and adapts naturally to the online workload.
We substitute \(\lambda\) to give the final scheduling rule:
\[
S^{*}
=
\arg\max_{S \in \mathcal{S}}
\text{Score}(S),
\quad
\text{Score}(S) =
\underbrace{J_{\text{gain}}(S)}_{\text{Throughput Gain}}
-
\underbrace{{|\mathcal{R}|}/{\bar{T}_{\text{exec}}^2} \cdot O_{\text{total}}(S)}_{\text{Cost Penalization}}
\]

\begin{algorithm}[tb]
   \caption{Flow-Adaptive Scheduling}
   \label{alg:cats}
\begin{algorithmic}[1]
   \STATE {\bfseries Input:} Request set $R$, System average execution time $\bar{T}$, Workload profiles $\Phi$
   \STATE {\bfseries Output:} Optimal decomposition size $s^*$, execution batch $Q_{\text{exec}}$, remain tasks $Q_{\text{remain}}$
   
   \STATE $s_{min}, s_{max} \leftarrow \text{get\_search\_space}(R)$
   \STATE $best\_score \leftarrow -\infty$
   \STATE $s^* \leftarrow s_{min}$
   
   \FOR{$s$ from $s_{min}$ to $s_{max}$}
       
       \STATE $exec\_time, overhead \leftarrow \text{estimate\_latency}(\Phi, R, s)$
       \STATE $total\_time \leftarrow exec\_time + overhead$

       \STATE $gain \leftarrow \sum_{r \in R} \min(1, s / r.workload)$
       
       \STATE $sensitivity \leftarrow |R| / \bar{T}^2$
       \STATE $waste \leftarrow \sum_{r \in R} (\lceil r.workload / s \rceil \times overhead)$
       
       \STATE $score \leftarrow (gain / total\_time) - (sensitivity \times waste)$
       
       \IF{$score > best\_score$}
           \STATE $best\_score \leftarrow score$
           \STATE $s^* \leftarrow s$
       \ENDIF
   \ENDFOR

   \STATE $Q_{\text{exec}}, Q_{\text{remain}} \leftarrow \text{construct}(R, s^*)$
   
   \STATE \bfseries return $s^*$, $Q_{\text{exec}}$, $Q_{\text{remain}}$
\end{algorithmic}
\end{algorithm}

\textbf{Online Scheduling Procedure.}
Algorithm~\ref{alg:cats} summarizes flow-adaptive scheduling. In each scheduling round, the scheduler receives requests from the queue and searches decomposition size within the valid search space (Lines~6). For each candidate size $S$, the scheduler estimates the execution time and decomposition overhead from the offline workload profiles (Lines~7), computes the throughput gain and cost penalization (Lines~10--12), and evaluates the scheduling score (Lines~13). The scheduler then greedily selects the decomposition size with the highest score (Lines~14--15).

After selecting the optimal decomposition size, the scheduler constructs the corresponding execution batch and remaining tasks (Lines~19). It then dispatches the batch to the execution engine and returns remaining tasks to the scheduler for subsequent scheduling rounds. In this way, the scheduler guarantees the gains of flow decomposition by selecting a decomposition size that reduces request blocking while controlling the overhead.


\section{Evaluation}
\label{sec:evaluation}
We evaluate \textsc{AgentWeave} on four representative agent applications introduced in Section~\ref {motivation:workflow}. For brevity, we provide detailed implementation and experimental setup in  Appendix~\ref{system implementation} and  Appendix~\ref{experimental setup}. 



\textbf{Baselines.}
We compare \textsc{AgentWeave} with two disaggregated architecture baselines that represent the optimization directions discussed in Section~\ref{disaggregated_inefficiency}.  
\textsc{Ayo}~\cite{ayo} optimizes agent workflow by splitting the modules into fine-grained task primitives and overlapping their execution to shorten the latency. 
\textsc{Autellix}~\cite{autellix} optimizes LLM serving by treating an agent program as the scheduling unit and prioritizing LLM calls using program-level execution context. 




\begin{figure}[t]
  \centering
  \includegraphics[width=\textwidth]{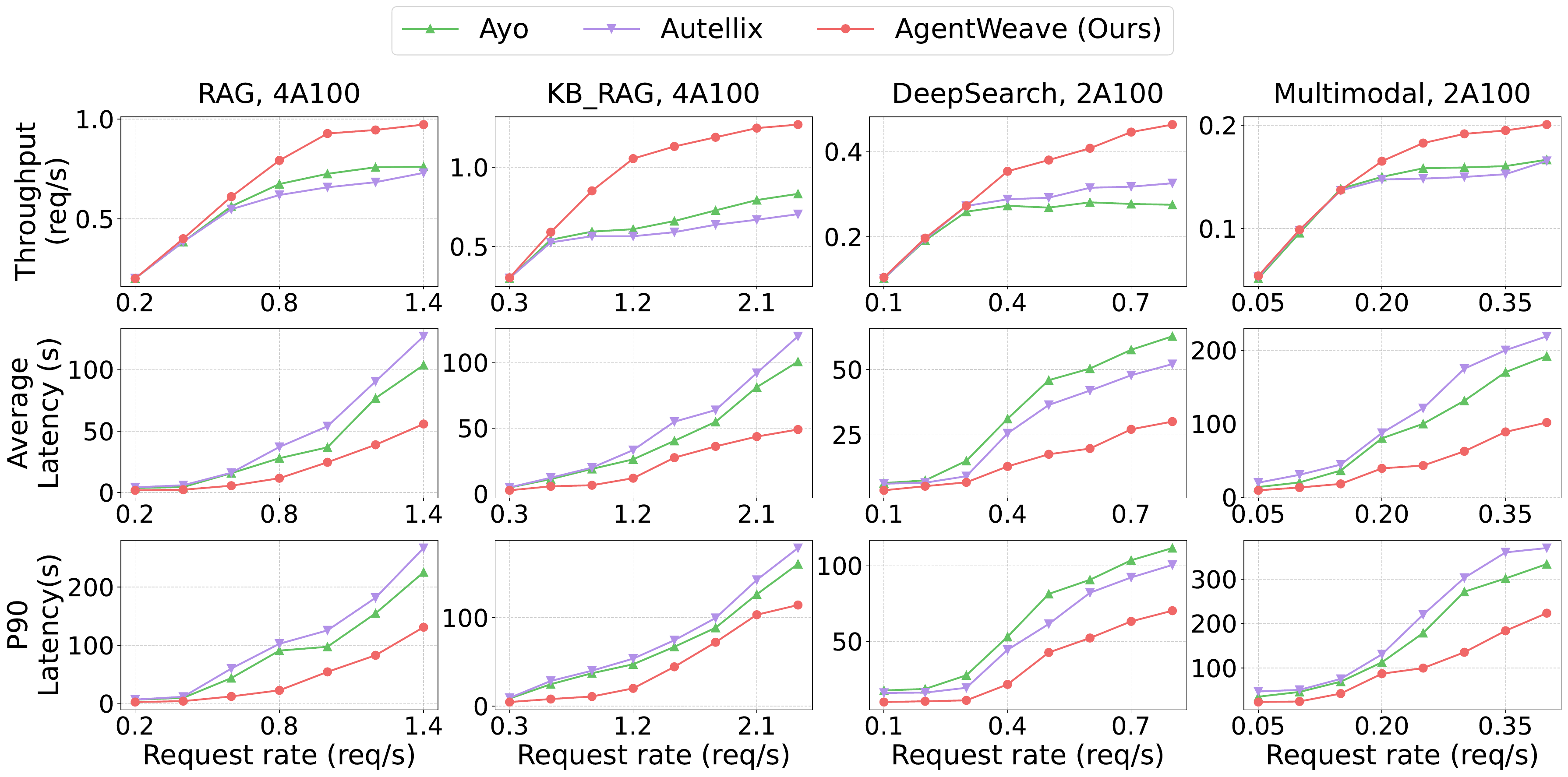} 
  \caption{End-to-end performance on four agent applications under varying request rates.}
  \label{fig:combined_performance}
  \vspace{-5pt}
\end{figure}




\subsection{Performance}
\label{sec:performance}

\textbf{Throughput.}
The first row of Figure~\ref{fig:combined_performance} shows that \textsc{AgentWeave} sustains higher throughput as input rate increases, improving over the baseline by 1.33$\times$ on RAG, 1.81$\times$ on KB-RAG, 1.68$\times$ on DeepSearch, and 1.28$\times$ on Multimodal. 
At low request rate, all systems perform similarly because the servers are not saturated. 
As rate increases, the disaggregated baselines saturate earlier: Autellix performs better on LLM-dominated DeepSearch agents, while Ayo performs better on other agents, but both are limited by pipeline bubbles. 
\textsc{AgentWeave} keeps devices busy and reduces request blocking, thereby sustaining higher throughput near saturation.

\textbf{Average latency.}
The second row of Figure~\ref{fig:combined_performance} shows that \textsc{AgentWeave} consistently reduces average latency, improving over the baseline by 2.28$\times$ on RAG, 2.45$\times$ on KB-RAG, 2.08$\times$ on DeepSearch, and 2.15$\times$ on Multimodal. 
At low input rates, queueing is limited, and the improvement mainly comes from aggregated execution, which shortens module execution. 
At higher input rates, \textsc{AgentWeave} further reduces blocking with flow decomposition, so latency grows more slowly.

\textbf{P90 latency.}
The third row of Figure~\ref{fig:combined_performance} shows that \textsc{AgentWeave} reduces tail latency over the baseline by 2.04$\times$ on RAG, 1.56$\times$ on KB-RAG, 1.59$\times$ on DeepSearch, and 1.66$\times$ on Multimodal. 
Unlike average latency, the P90 improvement mainly comes from the aggregated architecture. Tail requests are typically long requests themselves, so decomposition does not directly shorten their execution. However, the aggregated architecture still reduces the latency of these long requests by utilizing all devices to process each module.

\begin{figure*}[!h]
  \centering
  \includegraphics[width=0.83\textwidth]{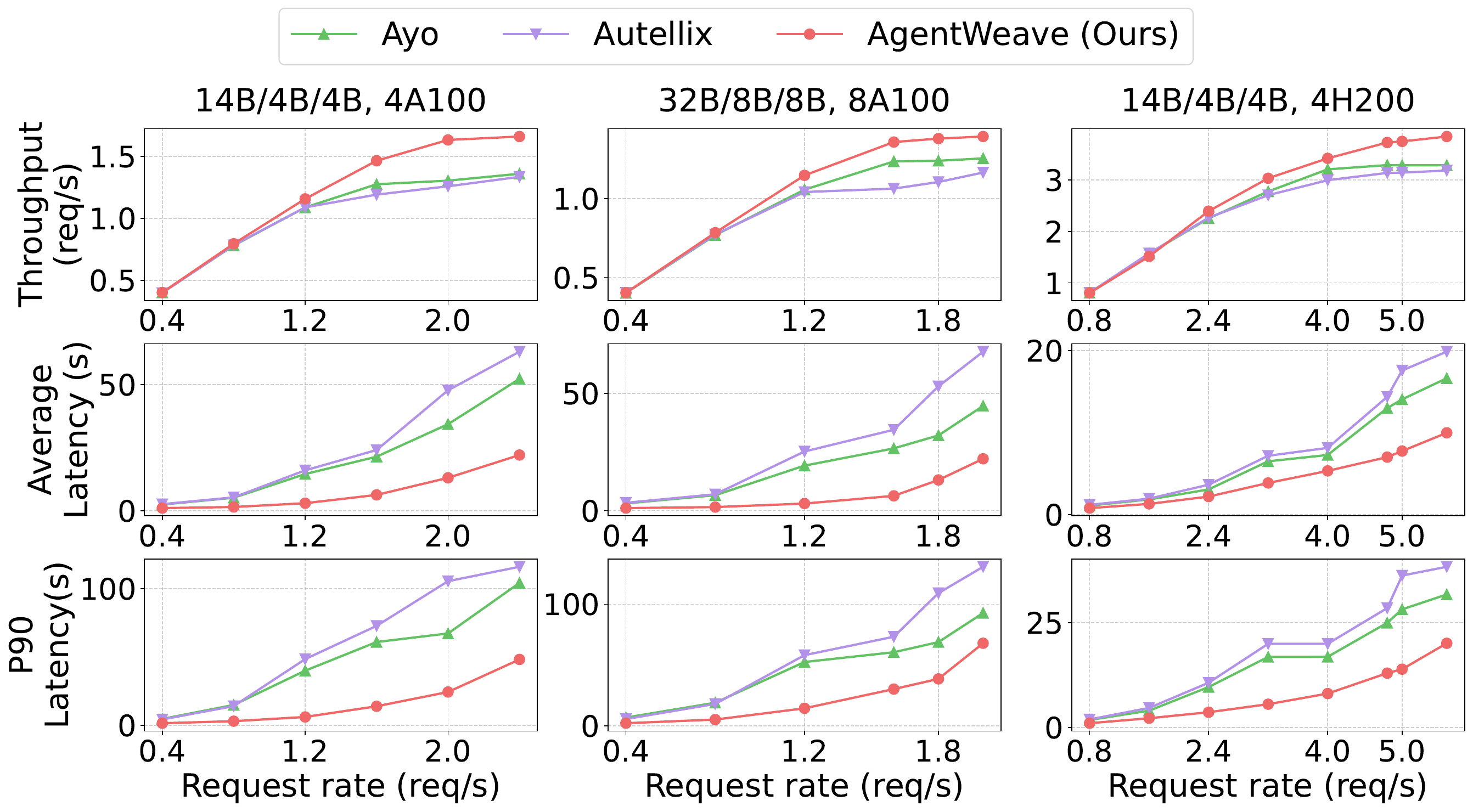} 
  \caption{End-to-end RAG performance under different model and hardware configurations. Model sizes are reported as LLM/Embedding/Reranker.}
  \label{fig:scale_performance}
\end{figure*}

\textbf{Different Configurations.}
\label{sec:hardware}
We further evaluate \textsc{AgentWeave} on RAG under different configurations in Figure~\ref{fig:scale_performance}.
\textsc{AgentWeave} improves throughput by 1.24$\times$, 1.20$\times$, and 1.21$\times$, reduces average latency by 2.85$\times$, 3.07$\times$, and 1.99$\times$, and reduces P90 latency by 2.41$\times$, 1.92$\times$, and 1.91$\times$, respectively.
The advantage becomes more pronounced under higher request rates, where baselines suffer from pipeline bubbles, and \textsc{AgentWeave} continues to reduce blocking in aggregated architecture. 
These gains show that \textsc{AgentWeave} remains effective when scaling to larger models, more devices, and across different GPU platforms.


\begin{figure*}[ht]
  \centering

  \begin{minipage}[t]{0.25\textwidth}
    \centering
    \includegraphics[width=\linewidth]{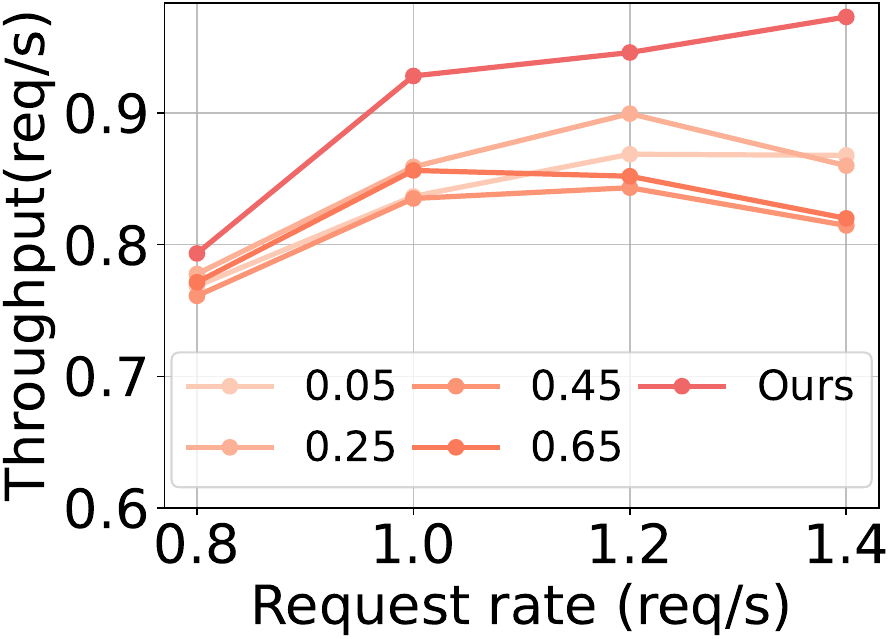}
    \captionof{figure}{Throughput for different trade-off parameter $\lambda$ in scheduling. }
    \label{fig:sensitivity}
  \end{minipage}
  \hfill
  \begin{minipage}[t]{0.73\textwidth}
    \centering
    \includegraphics[width=\linewidth]{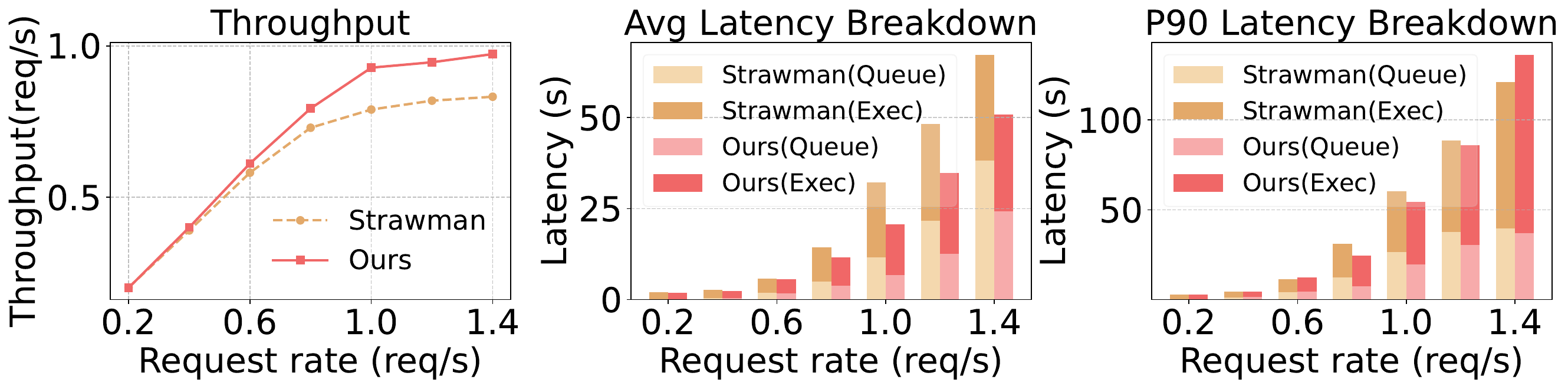}
    \captionof{figure}{Ablation studies of \textsc{AgentWeave} on RAG against the naive aggregated architecture. The figure reports end-to-end throughput, together with the breakdown of average and P90 latency.}
    \label{fig:ablation}
  \end{minipage}

\end{figure*}

\subsection{Ablation Studies and Sensitivity Analysis}

\textbf{Sensitivity Analysis.}
\label{sec:sensitivity}
We evaluate the effect of the trade-off parameter $\lambda$ on RAG in Figure~\ref{fig:sensitivity}.
We compare our $\lambda$ with static baselines: $\lambda=0.25$ selected by offline profiling, and $\lambda=0.05$, $0.45$, and $0.65$ covering different preferences over throughput gain and decomposition overhead. 
Our $\lambda$ achieves the highest throughput, improving over static settings by 1.13$\times$.
The advantage of our $\lambda$ comes from dynamically adapting to the current queue state.
Static values fit only limited regimes: $\lambda=0.25$ performs well in most cases but becomes suboptimal at QPS $=1.4$, while larger values over-penalize and suppress useful decomposition under high request rate.

\textbf{Ablation Studies.}
\label{sec:ablation}
We ablate \textsc{AgentWeave} against a naive aggregated architecture strawman on RAG in Figure~\ref{fig:ablation}. 
Both perform similarly at low request rates, as the servers are not saturated. At high request rates, \textsc{AgentWeave} reduces request blocking, improving throughput by 1.19$\times$ and reducing average queueing and execution latency by 1.57$\times$ and 1.48$\times$, respectively.
This confirms the gains from reducing request blocking.
Although decomposition adds overhead to long requests, flow-adaptive scheduling controls the decomposition size and keeps P90 latency comparable. 
We provide ablation results for the other three agent applications in Appendix~\ref{other ablation studies}.
Overall, \textsc{AgentWeave} improves throughput and average latency while limiting tail-latency overhead.

\subsection{Accuracy}
\label{sec:accuracy}


\textsc{AgentWeave} improves serving efficiency with flow decomposition while preserving the semantics of the original agent workflow. 
Following the setup in Section~\ref{sec:performance}, accuracy remains unchanged before and after applying \textsc{AgentWeave}: RAG stays at $0.8470$ on TriviaQA~\cite{triviaqa}, KB-RAG at $0.8170$ on TriviaQA~\cite{triviaqa}, and DeepSearch at $0.3097$ on GAIA~\cite{gaia}. 
This shows that \textsc{AgentWeave} preserves output quality while improving serving performance.


\section{Conclusion}
This paper presents \textsc{AgentWeave}, an efficient distributed serving system for agent applications. 
\textsc{AgentWeave} adopts an aggregated architecture and addresses its key challenge, request blocking, through decomposition.
It exposes flow as the decomposition unit and deploys flow decomposition to safely decompose a request.
It also deploys a flow-adaptive scheduler to balance decomposition gains against overhead. 
Experiments on four representative agents show that \textsc{AgentWeave} achieves better serving performance over existing baselines while preserving accuracy.
We discuss limitations and future directions in Appendix~\ref{sec:limitation}.



{
\small
\bibliographystyle{plainnat}
\bibliography{example_paper}
}


\clearpage
\appendix

\section{Agent Applications}
\label{sec:agent_primer}

This appendix introduces the four agent applications considered in this paper, covering their use cases and how they execute as workflows involving multiple LLM or tool invocations~\cite{gta}.

\subsection{RAG Agent}

Retrieval-augmented generation (RAG)~\cite{rag_2020,rag_2023} in Figure~\ref{fig:workflow}(a) is one of the most widely used agents for accurate answers generation with retrieved knowledge.
A standalone LLM answers from its parametric knowledge and the prompt context, which can be insufficient~\cite{rag_2023,webgpt} when the answer depends on newly added documents, private files, domain-specific corpora, or information that is too long to directly place in the prompt. 
A RAG agent addresses this limitation by coupling the LLM with retrieval tools. 
Instead of asking the LLM to answer directly, the agent first identifies relevant evidence and then asks the LLM to synthesize an answer conditioned on that evidence.

Given a user query and a set of documents, the RAG agent executes a workflow that combines LLM calls with tool calls. 
It first invokes an LLM to expand or rewrite the query~\cite{query_rewrite_v1, query_rewrite_v2}, producing a clearer search intent for downstream retrieval. 
The agent then calls an embedding tool to encode the documents and update the vector index. 
Using the rewritten query, it invokes a retrieval tool to search the index and obtain candidate documents. 
Since the retrieved candidates may be noisy or only partially relevant, the agent further invokes a reranking tool to rescore them and select the most relevant evidence. 
Finally, the agent invokes the LLM again to synthesize the answer from the reranked documents~\cite{rankrag}.

Modern RAG extends retrieval-augmented generation into a multi-stage agent workflow. 
It combines query expansion, document indexing, vector retrieval, reranking, and answer synthesis to improve knowledge-grounded response generation. 
Advanced RAG therefore represents a common agent workflow that relies on multiple LLM and tool invocations for knowledge-intensive tasks.

\subsection{Knowledge-Based RAG Agent}

Knowledge-based RAG~\cite{rag_2023,hedrarag} in Figure~\ref{fig:workflow}(b) is an agent for answering questions over a persistent knowledge source~\cite{kb}. 
It follows the same general principle as RAG: the LLM should answer using retrieved evidence rather than relying only on its internal knowledge. 
Different from advanced RAG, the documents are processed and indexed offline into a local knowledge base, which may contain enterprise documents, product manuals, internal wikis, troubleshooting guides, code repositories, or other domain-specific resources~\cite{kb,amazon_q_connectors,google_cloud_search_connectors}. 
This setting is widely used when user queries target private, domain-specific, or frequently updated corpora.

Given a user query, the knowledge-based RAG agent executes a workflow that combines LLM calls with retrieval tools. 
It first invokes an LLM to expand or rewrite the query, producing a clearer search intent for the knowledge base. 
The agent then calls a retrieval tool to search the pre-built database and obtain candidate documents or passages. 
Since the retrieved candidates may still contain irrelevant or weakly related evidence, the agent further invokes a reranking tool to rescore them and select the most relevant passages. 
Finally, the agent invokes the LLM again to synthesize the answer from the reranked evidence.

Knowledge-based RAG removes the online document encoding stage and focuses on query expansion, knowledge-base retrieval, reranking, and answer synthesis over a persistent corpus. 
This makes it a practical agent workflow for domain-specific question answering, where LLM reasoning is coupled with retrieval from private or enterprise knowledge sources.

\subsection{DeepSearch Multi-Agent}

DeepSearch multi-agent~\cite{deepsearch,manusearch} in Figure~\ref{fig:workflow}(c) is an agent for complex information-seeking tasks. 
Unlike RAG-style agents, which usually retrieve evidence from a fixed corpus or index, DeepSearch actively explores external sources through multiple coordinated agents. 
It is designed for questions that require broad investigation, where the answer may depend on collecting evidence from different directions and synthesizing them into a structured report.

Given a user query, the DeepSearch multi-agent executes a workflow that combines LLM agents with search subagent tools~\cite{camel}. 
It first invokes a lead LLM agent to analyze the query and decompose it into multiple subquestions or research directions. 
The lead agent then dispatches these subquestions to multiple search subagents. 
Each subagent invokes search tools, reads retrieved pages or documents, refines its search when needed, and summarizes the evidence for its assigned direction. 
After the subagents finish, a consolidation agent aggregates their findings, removes redundant information, resolves inconsistent evidence, and generates the final response with supporting citations.

DeepSearch organizes complex information seeking as a multi-agent research process. 
By decomposing a query into multiple directions, collecting evidence in parallel, and consolidating intermediate findings, it supports open-ended questions that require broader exploration than standard retrieval-based workflows.

\subsection{Multimodal Generation Agent}

Multimodal generation~\cite{rpg,llmgrounded, multimodal_serving} in Figure~\ref{fig:workflow}(d) is an agent for generating images from text prompts. 
A diffusion model~\cite{sd3} can synthesize an image directly from a prompt, but short user prompts often underspecify visual details such as object attributes, spatial relations, background, lighting, or style. 
A multimodal generation agent addresses this limitation by using LLM calls to refine and structure the prompt before invoking the image generator. 
This improves prompt understanding, scene composition, and controllability.

Given a user prompt, the multimodal generation agent executes a workflow that combines LLM calls with a generative tool. 
It first invokes an LLM-based recaptioning module to rewrite the prompt into a richer and more explicit description. 
The agent then invokes an LLM-based planning module to decompose the refined description into structured scene plans, such as objects, regions, layouts, or visual constraints. 
Finally, the agent calls a diffusion-based image generation tool to synthesize the final image from the planned description.

Multimodal generation extends agent workflows beyond text-only question answering. 
It uses LLM-based recaptioning and planning to prepare structured inputs for a diffusion-based generator, connecting language reasoning with specialized visual generation tools.




\subsection{Common Properties}

Although the four applications target different tasks, they share several properties that distinguish agent serving from serving standalone LLM requests~\cite{klotski,d2moe,vllm,sglang}. 
First, they execute as module workflows. 
The final response is produced through multiple LLM calls and tool invocations, such as retrieval, reranking, search, database access, or image generation. 
Second, these modules are connected by data dependencies. 
Each module consumes intermediate data produced by earlier modules, so the workflow can only progress when the required data becomes available. 
Third, requests can have substantially different execution costs. 
Different requests carry different documents, retrieved candidates, subquestions, search results, or scene plans, which change the amount of work required by each module and lead to large differences in execution time.

These properties make agent serving a workflow-level systems problem. 
Optimizing a single LLM call or an individual tool is insufficient, because overall performance also depends on how intermediate data moves through the workflow and how different modules are scheduled together. 
Therefore, an agent serving system should optimize the execution of dependency-constrained module workflows rather than treating each module invocation as an isolated request.

\section{Other Decomposition}
\label{other_decomposition}

\begin{figure}[!h]
  \centering
  \includegraphics[width=\textwidth]{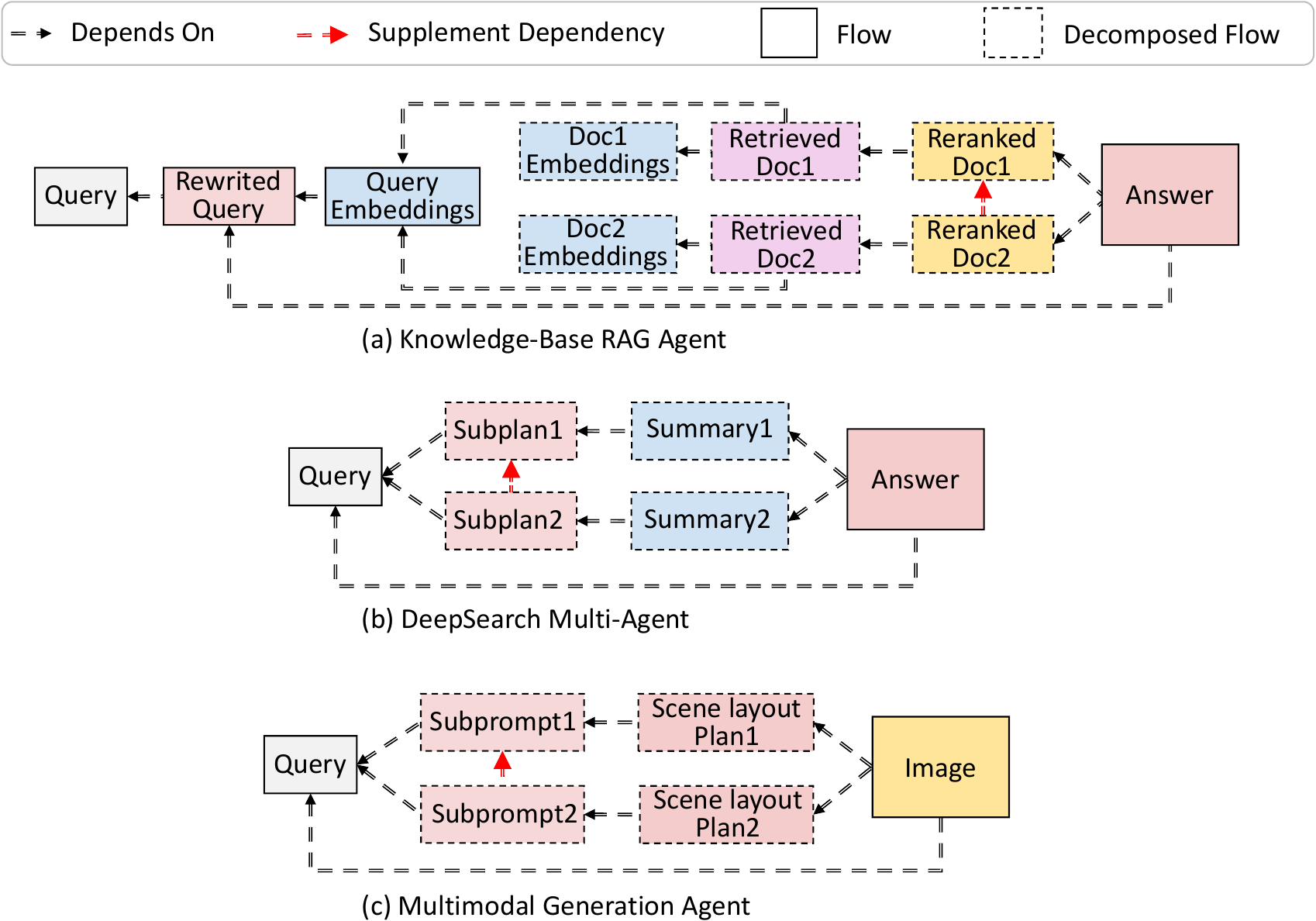} 
  \caption{Flow decomposition on the KB-RAG, deepsearch and multimodal agent in Figure~\ref{fig:workflow}(b)(c)(d).}
  \label{fig:appendix_decomposition}
\end{figure}

Figure~\ref{fig:appendix_decomposition} illustrates flow decomposition for the other three agent applications in Figure~\ref{fig:workflow}. 
These decompositions follow the same construction rule as Section~\ref{flow_abstraction} and Section~\ref{decomposition}: we rewrite the module workflow into a flow dependency graph, split decomposable flows along valid dependencies, decompose conditional decomposable flows, and insert supplement dependencies to satisfy the complete-input requirement.

\textbf{Knowledge-Base RAG Agent.}
The KB-RAG agent in Figure~\ref{fig:workflow}(b) follows the retrieval, reranking, and synthesis stages of RAG, but retrieves documents from a pre-built knowledge base instead of indexing newly provided documents.
The rewritten query and query embedding are kept intact, since they describe a single query intent and cannot be decomposed into independent subflows.
The retrieved documents and document embeddings are decomposable, thus split into independent subflows.
Reranked documents are conditionally decomposable, since reranking needs the complete retrieved documents to select global top-$k$ documents. We decompose them and insert supplement dependencies to satisfy the complete-input requirement.

\textbf{DeepSearch Multi-Agent.}
The DeepSearch workflow in Figure~\ref{fig:workflow}(c) contains query decomposition, subagent search, summary generation, and result consolidation.
The original query is still kept intact.
The summaries are decomposable, because they can be split to independent summary on different plan, and each is a correct subset of the full summaries.
The plan is conditionally decomposable, because each subplan describes a separate search target, but the plan still requires complete input.
We decompose it and insert supplement dependencies, which corresponds to decomposing the output of LLM generation.

\textbf{Multimodal Generation Agent.}
The multimodal generation workflow in Figure~\ref{fig:workflow}(d) contains query recaption, scene planning, and image generation.
The original query is kept intact, since it describes the user intent.
The scene-layout plans are decomposable, because they can be split into independent plans, and each is a correct subset of the full scene-layout flow.
The subprompts are conditionally decomposable, because each subprompt describes an independent visual generation target, but the subprompt flow still requires the complete query as input.
Also, we decompose it and insert supplement dependencies.

\section{System Implementation}
\label{system implementation}

\textsc{AgentWeave} is implemented as a distributed serving system for agent workflows, with roughly 8.9K lines of Python code, based on FastAPI~\cite{fastapi} for the control plane, Ray~\cite{ray} for distributed scheduling and execution, SGLang~\cite{sglang} for LLM generation, embedding, and reranking, FAISS~\cite{faiss} for retrieval, xDiT~\cite{xdit} for diffusion-based generation.

\textbf{Frontend.}
The frontend is an asynchronous FastAPI service that exposes a stateless request interface, while the runtime tracks each request as a long-lived stateful flow.
After admission, a request enters an FCFS queue that stores remaining subflows, intermediate state, and completion status.
This separates the user-facing API from decomposition state, allowing \textsc{AgentWeave} to pause and resume requests across scheduling rounds without changing application semantics.

\textbf{Execution engine.} \textsc{AgentWeave} wraps SGLang's asynchronous engine interface as the LLM executor, with continuous batching~\cite{orca}, chunked prefill~\cite{sarathi-serve} and prefix caching~\cite{sglang} techniques. \textsc{AgentWeave} keeps decomposition and scheduling decisions outside the LLM engine.

\textbf{Tool backends.}
Beyond the LLM executor, \textsc{AgentWeave} implements embedding, reranking, retrieval, web search, and multimodal generation as resident tool services, avoiding repeated initialization during decomposed execution.

\textbf{Scheduler.}
The scheduler is an event-driven loop over active request states.
It is independent of tool internals and agent-specific module code. 
In each scheduling round, it uses profiled data to estimate execution time and scheduling overhead, selects the decomposition size, constructs the execution batch, and returns remaining tasks to the queue.


\textbf{Agent executors.}
Each application uses a common executor interface to define its flow semantics, including which flows can be decomposed, which context remains intact, and which intermediate results are carried as supplement dependencies.
Our prototype implements this interface for RAG, KB-RAG, DeepSearch, and multimodal generation.
Thus, application-specific decomposition rules remain inside each executor, while the frontend, scheduler, LLM executor, and tool backends are shared across agents.

\section{Experimental Setup}
\label{experimental setup}
We evaluate \textsc{AgentWeave} on four representative agent applications: RAG, KB-RAG, DeepSearch, and multimodal generation. 
For all experiments, we use the same hardware for \textsc{AgentWeave} and the baselines. 
The disaggregated baselines place different modules on separate GPU groups, while \textsc{AgentWeave} adopts the aggregated architecture and co-deploys modules across all available GPUs.

\textbf{RAG.}
We evaluate RAG on TriviaQA~\cite{triviaqa} with two model configurations. 
The 14B/4B/4B configuration uses Qwen3-14B~\cite{qwen3} as the LLM, Qwen3-Embedding-4B~\cite{qwen3-embd} as the embedding model, and Qwen3-Reranker-4B~\cite{qwen3-embd} as the reranker. 
The 32B/8B/8B configuration uses Qwen3-32B~\cite{qwen3}, Qwen3-Embedding-8B~\cite{qwen3-embd}, and Qwen3-Reranker-8B~\cite{qwen3-embd}. 
We evaluate 32B/8B/8B on 4 A100 GPUs~\cite{a100} in Figure~\ref{fig:combined_performance}, 
14B/4B/4B on 4 A100 GPUs, 32B/8B/8B on 8 A100 GPUs, and 14B/4B/4B on 4 H200 GPUs~\cite{h200} in Figure~\ref{fig:scale_performance}.
For the 4-GPU settings, the disaggregated architecture deploys the embedding model and reranker on one GPU each, and deploys the LLM with tensor parallelism~\cite{megatron} across the remaining two GPUs. 
For the 8-GPU setting, the disaggregated architecture deploys the embedding model and reranker with tensor parallelism across two GPUs each, and deploys the LLM with tensor parallelism across remaining four GPUs. 
For \textsc{AgentWeave}, all models are deployed with tensor parallelism across all available GPUs.

\textbf{KB-RAG.}
KB-RAG follows the RAG workflow but retrieves candidate knowledge from a pre-built knowledge base instead of indexing documents at request time. 
We evaluate KB-RAG on TriviaQA using Qwen3-32B, Qwen3-Embedding-8B, and Qwen3-Reranker-8B on 4 A100 GPUs in Figure~\ref{fig:combined_performance}. 
For the disaggregated architecture, the embedding model and reranker are deployed on one GPU each, and the LLM is deployed with tensor parallelism across two GPUs. 
For \textsc{AgentWeave}, all models are deployed with tensor parallelism across all four GPUs.

\textbf{DeepSearch.}
We evaluate DeepSearch on GAIA~\cite{gaia}. 
We use Qwen3-14B as the main LLM and Qwen3-4B~\cite{qwen3} as the subagent SLM in Figure~\ref{fig:combined_performance}. 
Experiments are conducted on 2 A100 GPUs. 
For the disaggregated architecture, the main LLM and subagent model are deployed on one GPU each. 
For \textsc{AgentWeave}, both models are deployed with tensor parallelism across all GPUs.

\textbf{Multimodal Generation.}
We evaluate multimodal generation on T2I-CompBench~\cite{t2i}. 
We use Qwen3-14B for recaptioning and planning, and Stable Diffusion 3~\cite{sd3} as the diffusion generator in Figure~\ref{fig:combined_performance}. 
Experiments are conducted on 2 A100 GPUs. 
For the disaggregated architecture, the LLM and diffusion generator are deployed on one GPU each. 
For \textsc{AgentWeave}, the LLM is deployed with tensor parallelism across all GPUs, and the diffusion generator with classifier-free guidance parallelism~\cite{classifier} and PipeFusion-style parallelism~\cite{xdit}.

\section{Ablation Studies}
\label{other ablation studies}

\begin{figure*}[!h]
  \centering
  \includegraphics[width=\textwidth]{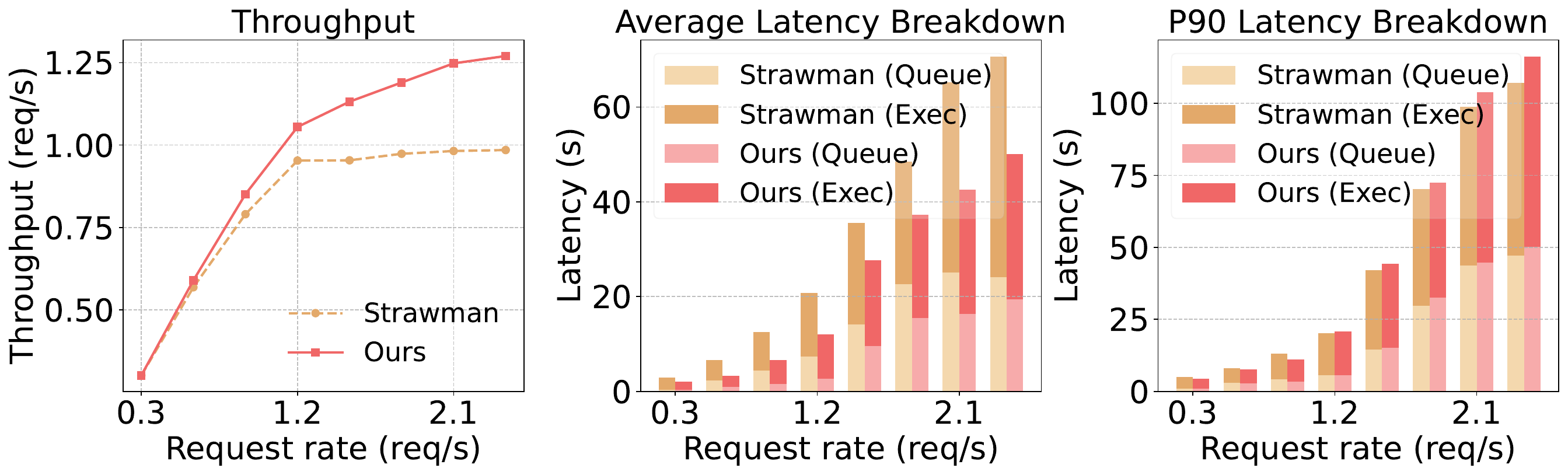} 
  \caption{Ablation studies of \textsc{AgentWeave} on Knowledge-Based RAG Agent against the naive aggregated architecture. The figure reports end-to-end throughput, together with the breakdown of average and P90 latency.}
  \label{fig:kb_ablation}
\end{figure*}

\begin{figure*}[!h]
  \centering
  \includegraphics[width=\textwidth]{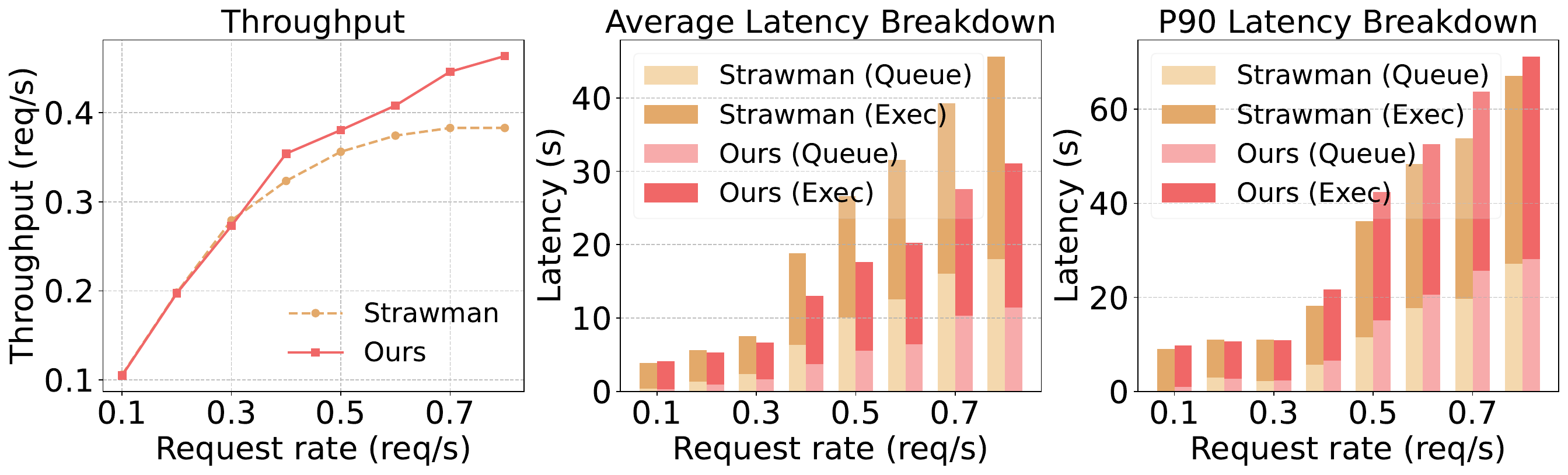} 
  \caption{Ablation studies of \textsc{AgentWeave} on DeepSearch Multi-Agent against the naive aggregated architecture. The figure reports end-to-end throughput, together with the breakdown of average and P90 latency.}
  \label{fig:dr_ablation}
\end{figure*}

\begin{figure*}[!h]
  \centering
  \includegraphics[width=\textwidth]{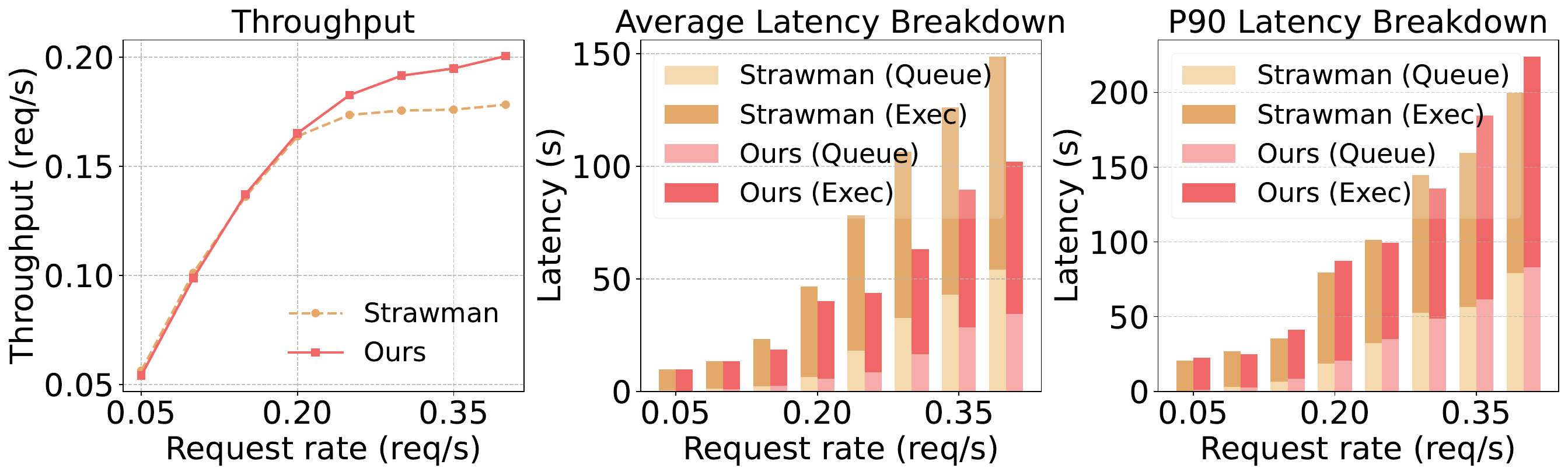} 
  \caption{Ablation studies of \textsc{AgentWeave} on Multimodal Generation Agent against the naive aggregated architecture. The figure reports end-to-end throughput, together with the breakdown of average and P90 latency.}
  \label{fig:rpg_ablation}
\end{figure*}

\section{Limitations and Future Work}
\label{sec:limitation}

\textsc{AgentWeave} has several limitations. 
First, it currently requires each agent application to specify its flow semantics, including decomposable flows and complete-input dependencies. 
Automatically deriving such semantics from general agent programs remains future work. 
Second, the flow-adaptive scheduler uses offline profiles to estimate execution time and decomposition overhead; online profile adaptation could further improve robustness under workload and hardware changes. 
Finally, our evaluation focuses on four representative agents. 
Extending \textsc{AgentWeave} to more dynamic production agents, external-tool failures, and multi-tenant deployments is an important future direction.


\clearpage
\section*{NeurIPS Paper Checklist}

\begin{enumerate}

\item {\bf Claims}
    \item[] Question: Do the main claims made in the abstract and introduction accurately reflect the paper's contributions and scope?
    \item[] Answer: \answerYes{} 
    \item[] Justification: The abstract and introduction state the paper's scope and main contributions: an aggregated architecture, flow decomposition, and a flow-adaptive scheduler for agent serving. The experimental claims are supported by the evaluation in Section~\ref{sec:evaluation}.
    \item[] Guidelines:
    \begin{itemize}
        \item The answer \answerNA{} means that the abstract and introduction do not include the claims made in the paper.
        \item The abstract and/or introduction should clearly state the claims made, including the contributions made in the paper and important assumptions and limitations. A \answerNo{} or \answerNA{} answer to this question will not be perceived well by the reviewers. 
        \item The claims made should match theoretical and experimental results, and reflect how much the results can be expected to generalize to other settings. 
        \item It is fine to include aspirational goals as motivation as long as it is clear that these goals are not attained by the paper. 
    \end{itemize}

\item {\bf Limitations}
    \item[] Question: Does the paper discuss the limitations of the work performed by the authors?
    \item[] Answer: \answerYes{} 
    \item[] Justification: We discuss the main limitations of \textsc{AgentWeave} and directions for future work in Appendix~\ref{sec:limitation}.
    \item[] Guidelines:
    \begin{itemize}
        \item The answer \answerNA{} means that the paper has no limitation while the answer \answerNo{} means that the paper has limitations, but those are not discussed in the paper. 
        \item The authors are encouraged to create a separate ``Limitations'' section in their paper.
        \item The paper should point out any strong assumptions and how robust the results are to violations of these assumptions (e.g., independence assumptions, noiseless settings, model well-specification, asymptotic approximations only holding locally). The authors should reflect on how these assumptions might be violated in practice and what the implications would be.
        \item The authors should reflect on the scope of the claims made, e.g., if the approach was only tested on a few datasets or with a few runs. In general, empirical results often depend on implicit assumptions, which should be articulated.
        \item The authors should reflect on the factors that influence the performance of the approach. For example, a facial recognition algorithm may perform poorly when image resolution is low or images are taken in low lighting. Or a speech-to-text system might not be used reliably to provide closed captions for online lectures because it fails to handle technical jargon.
        \item The authors should discuss the computational efficiency of the proposed algorithms and how they scale with dataset size.
        \item If applicable, the authors should discuss possible limitations of their approach to address problems of privacy and fairness.
        \item While the authors might fear that complete honesty about limitations might be used by reviewers as grounds for rejection, a worse outcome might be that reviewers discover limitations that aren't acknowledged in the paper. The authors should use their best judgment and recognize that individual actions in favor of transparency play an important role in developing norms that preserve the integrity of the community. Reviewers will be specifically instructed to not penalize honesty concerning limitations.
    \end{itemize}

\item {\bf Theory assumptions and proofs}
    \item[] Question: For each theoretical result, does the paper provide the full set of assumptions and a complete (and correct) proof?
    \item[] Answer: \answerNA{} 
    \item[] Justification: The paper does not present formal theoretical results.
    \item[] Guidelines:
    \begin{itemize}
        \item The answer \answerNA{} means that the paper does not include theoretical results. 
        \item All the theorems, formulas, and proofs in the paper should be numbered and cross-referenced.
        \item All assumptions should be clearly stated or referenced in the statement of any theorems.
        \item The proofs can either appear in the main paper or the supplemental material, but if they appear in the supplemental material, the authors are encouraged to provide a short proof sketch to provide intuition. 
        \item Inversely, any informal proof provided in the core of the paper should be complemented by formal proofs provided in appendix or supplemental material.
        \item Theorems and Lemmas that the proof relies upon should be properly referenced. 
    \end{itemize}

    \item {\bf Experimental result reproducibility}
    \item[] Question: Does the paper fully disclose all the information needed to reproduce the main experimental results of the paper to the extent that it affects the main claims and/or conclusions of the paper (regardless of whether the code and data are provided or not)?
    \item[] Answer: \answerYes{} 
    \item[] Justification: We provide implementation details and experimental settings in Appendix~\ref{system implementation} and Appendix~\ref{experimental setup}.
    \item[] Guidelines:
    \begin{itemize}
        \item The answer \answerNA{} means that the paper does not include experiments.
        \item If the paper includes experiments, a \answerNo{} answer to this question will not be perceived well by the reviewers: Making the paper reproducible is important, regardless of whether the code and data are provided or not.
        \item If the contribution is a dataset and\slash or model, the authors should describe the steps taken to make their results reproducible or verifiable. 
        \item Depending on the contribution, reproducibility can be accomplished in various ways. For example, if the contribution is a novel architecture, describing the architecture fully might suffice, or if the contribution is a specific model and empirical evaluation, it may be necessary to either make it possible for others to replicate the model with the same dataset, or provide access to the model. In general. releasing code and data is often one good way to accomplish this, but reproducibility can also be provided via detailed instructions for how to replicate the results, access to a hosted model (e.g., in the case of a large language model), releasing of a model checkpoint, or other means that are appropriate to the research performed.
        \item While NeurIPS does not require releasing code, the conference does require all submissions to provide some reasonable avenue for reproducibility, which may depend on the nature of the contribution. For example
        \begin{enumerate}
            \item If the contribution is primarily a new algorithm, the paper should make it clear how to reproduce that algorithm.
            \item If the contribution is primarily a new model architecture, the paper should describe the architecture clearly and fully.
            \item If the contribution is a new model (e.g., a large language model), then there should either be a way to access this model for reproducing the results or a way to reproduce the model (e.g., with an open-source dataset or instructions for how to construct the dataset).
            \item We recognize that reproducibility may be tricky in some cases, in which case authors are welcome to describe the particular way they provide for reproducibility. In the case of closed-source models, it may be that access to the model is limited in some way (e.g., to registered users), but it should be possible for other researchers to have some path to reproducing or verifying the results.
        \end{enumerate}
    \end{itemize}

\item {\bf Open access to data and code}
    \item[] Question: Does the paper provide open access to the data and code, with sufficient instructions to faithfully reproduce the main experimental results, as described in supplemental material?
    \item[] Answer: \answerNo{} 
    \item[] Justification: We do not release code at submission time, but provide implementation and setup details for reproducibility.
    \item[] Guidelines:
    \begin{itemize}
        \item The answer \answerNA{} means that paper does not include experiments requiring code.
        \item Please see the NeurIPS code and data submission guidelines (\url{https://neurips.cc/public/guides/CodeSubmissionPolicy}) for more details.
        \item While we encourage the release of code and data, we understand that this might not be possible, so \answerNo{} is an acceptable answer. Papers cannot be rejected simply for not including code, unless this is central to the contribution (e.g., for a new open-source benchmark).
        \item The instructions should contain the exact command and environment needed to run to reproduce the results. See the NeurIPS code and data submission guidelines (\url{https://neurips.cc/public/guides/CodeSubmissionPolicy}) for more details.
        \item The authors should provide instructions on data access and preparation, including how to access the raw data, preprocessed data, intermediate data, and generated data, etc.
        \item The authors should provide scripts to reproduce all experimental results for the new proposed method and baselines. If only a subset of experiments are reproducible, they should state which ones are omitted from the script and why.
        \item At submission time, to preserve anonymity, the authors should release anonymized versions (if applicable).
        \item Providing as much information as possible in supplemental material (appended to the paper) is recommended, but including URLs to data and code is permitted.
    \end{itemize}

\item {\bf Experimental setting/details}
    \item[] Question: Does the paper specify all the training and test details (e.g., data splits, hyperparameters, how they were chosen, type of optimizer) necessary to understand the results?
    \item[] Answer: \answerYes{} 
    \item[] Justification: We specify workloads, models, hardware, baselines, request rates, and metrics in Section~\ref{sec:evaluation} and Appendix~\ref{experimental setup}.
    \item[] Guidelines:
    \begin{itemize}
        \item The answer \answerNA{} means that the paper does not include experiments.
        \item The experimental setting should be presented in the core of the paper to a level of detail that is necessary to appreciate the results and make sense of them.
        \item The full details can be provided either with the code, in appendix, or as supplemental material.
    \end{itemize}

\item {\bf Experiment statistical significance}
    \item[] Question: Does the paper report error bars suitably and correctly defined or other appropriate information about the statistical significance of the experiments?
    \item[] Answer: \answerNo{} 
    \item[] Justification: We do not report error bars or significance tests.
    \item[] Guidelines:
    \begin{itemize}
        \item The answer \answerNA{} means that the paper does not include experiments.
        \item The authors should answer \answerYes{} if the results are accompanied by error bars, confidence intervals, or statistical significance tests, at least for the experiments that support the main claims of the paper.
        \item The factors of variability that the error bars are capturing should be clearly stated (for example, train/test split, initialization, random drawing of some parameter, or overall run with given experimental conditions).
        \item The method for calculating the error bars should be explained (closed form formula, call to a library function, bootstrap, etc.)
        \item The assumptions made should be given (e.g., Normally distributed errors).
        \item It should be clear whether the error bar is the standard deviation or the standard error of the mean.
        \item It is OK to report 1-sigma error bars, but one should state it. The authors should preferably report a 2-sigma error bar than state that they have a 96\% CI, if the hypothesis of Normality of errors is not verified.
        \item For asymmetric distributions, the authors should be careful not to show in tables or figures symmetric error bars that would yield results that are out of range (e.g., negative error rates).
        \item If error bars are reported in tables or plots, the authors should explain in the text how they were calculated and reference the corresponding figures or tables in the text.
    \end{itemize}

\item {\bf Experiments compute resources}
    \item[] Question: For each experiment, does the paper provide sufficient information on the computer resources (type of compute workers, memory, time of execution) needed to reproduce the experiments?
    \item[] Answer: \answerYes{} 
    \item[] Justification: We report the GPU configurations, model placements, and experimental setup in Appendix~\ref{experimental setup}.
    \item[] Guidelines:
    \begin{itemize}
        \item The answer \answerNA{} means that the paper does not include experiments.
        \item The paper should indicate the type of compute workers CPU or GPU, internal cluster, or cloud provider, including relevant memory and storage.
        \item The paper should provide the amount of compute required for each of the individual experimental runs as well as estimate the total compute. 
        \item The paper should disclose whether the full research project required more compute than the experiments reported in the paper (e.g., preliminary or failed experiments that didn't make it into the paper). 
    \end{itemize}
    
\item {\bf Code of ethics}
    \item[] Question: Does the research conducted in the paper conform, in every respect, with the NeurIPS Code of Ethics \url{https://neurips.cc/public/EthicsGuidelines}?
    \item[] Answer: \answerYes{} 
    \item[] Justification: The work uses public datasets, models, and tools, and does not involve human-subject experiments or sensitive data.
    \item[] Guidelines:
    \begin{itemize}
        \item The answer \answerNA{} means that the authors have not reviewed the NeurIPS Code of Ethics.
        \item If the authors answer \answerNo, they should explain the special circumstances that require a deviation from the Code of Ethics.
        \item The authors should make sure to preserve anonymity (e.g., if there is a special consideration due to laws or regulations in their jurisdiction).
    \end{itemize}

\item {\bf Broader impacts}
    \item[] Question: Does the paper discuss both potential positive societal impacts and negative societal impacts of the work performed?
    \item[] Answer: \answerYes{} 
    \item[] Justification: The work can improve the efficiency of agent serving.
    \item[] Guidelines:
    \begin{itemize}
        \item The answer \answerNA{} means that there is no societal impact of the work performed.
        \item If the authors answer \answerNA{} or \answerNo, they should explain why their work has no societal impact or why the paper does not address societal impact.
        \item Examples of negative societal impacts include potential malicious or unintended uses (e.g., disinformation, generating fake profiles, surveillance), fairness considerations (e.g., deployment of technologies that could make decisions that unfairly impact specific groups), privacy considerations, and security considerations.
        \item The conference expects that many papers will be foundational research and not tied to particular applications, let alone deployments. However, if there is a direct path to any negative applications, the authors should point it out. For example, it is legitimate to point out that an improvement in the quality of generative models could be used to generate Deepfakes for disinformation. On the other hand, it is not needed to point out that a generic algorithm for optimizing neural networks could enable people to train models that generate Deepfakes faster.
        \item The authors should consider possible harms that could arise when the technology is being used as intended and functioning correctly, harms that could arise when the technology is being used as intended but gives incorrect results, and harms following from (intentional or unintentional) misuse of the technology.
        \item If there are negative societal impacts, the authors could also discuss possible mitigation strategies (e.g., gated release of models, providing defenses in addition to attacks, mechanisms for monitoring misuse, mechanisms to monitor how a system learns from feedback over time, improving the efficiency and accessibility of ML).
    \end{itemize}
    
\item {\bf Safeguards}
    \item[] Question: Does the paper describe safeguards that have been put in place for responsible release of data or models that have a high risk for misuse (e.g., pre-trained language models, image generators, or scraped datasets)?
    \item[] Answer: \answerNA{} 
    \item[] Justification: The paper does not release new models, generators, or scraped datasets with high misuse risk.
    \item[] Guidelines:
    \begin{itemize}
        \item The answer \answerNA{} means that the paper poses no such risks.
        \item Released models that have a high risk for misuse or dual-use should be released with necessary safeguards to allow for controlled use of the model, for example by requiring that users adhere to usage guidelines or restrictions to access the model or implementing safety filters. 
        \item Datasets that have been scraped from the Internet could pose safety risks. The authors should describe how they avoided releasing unsafe images.
        \item We recognize that providing effective safeguards is challenging, and many papers do not require this, but we encourage authors to take this into account and make a best faith effort.
    \end{itemize}

\item {\bf Licenses for existing assets}
    \item[] Question: Are the creators or original owners of assets (e.g., code, data, models), used in the paper, properly credited and are the license and terms of use explicitly mentioned and properly respected?
    \item[] Answer: \answerYes{} 
    \item[] Justification: They are open-sourced.
    \item[] Guidelines:
    \begin{itemize}
        \item The answer \answerNA{} means that the paper does not use existing assets.
        \item The authors should cite the original paper that produced the code package or dataset.
        \item The authors should state which version of the asset is used and, if possible, include a URL.
        \item The name of the license (e.g., CC-BY 4.0) should be included for each asset.
        \item For scraped data from a particular source (e.g., website), the copyright and terms of service of that source should be provided.
        \item If assets are released, the license, copyright information, and terms of use in the package should be provided. For popular datasets, \url{paperswithcode.com/datasets} has curated licenses for some datasets. Their licensing guide can help determine the license of a dataset.
        \item For existing datasets that are re-packaged, both the original license and the license of the derived asset (if it has changed) should be provided.
        \item If this information is not available online, the authors are encouraged to reach out to the asset's creators.
    \end{itemize}

\item {\bf New assets}
    \item[] Question: Are new assets introduced in the paper well documented and is the documentation provided alongside the assets?
    \item[] Answer: \answerNA{} 
    \item[] Justification: The paper does not release new datasets or models.
    \item[] Guidelines:
    \begin{itemize}
        \item The answer \answerNA{} means that the paper does not release new assets.
        \item Researchers should communicate the details of the dataset\slash code\slash model as part of their submissions via structured templates. This includes details about training, license, limitations, etc. 
        \item The paper should discuss whether and how consent was obtained from people whose asset is used.
        \item At submission time, remember to anonymize your assets (if applicable). You can either create an anonymized URL or include an anonymized zip file.
    \end{itemize}

\item {\bf Crowdsourcing and research with human subjects}
    \item[] Question: For crowdsourcing experiments and research with human subjects, does the paper include the full text of instructions given to participants and screenshots, if applicable, as well as details about compensation (if any)? 
    \item[] Answer: \answerNA{} 
    \item[] Justification: The paper does not involve crowdsourcing or human-subject studies.
    \item[] Guidelines:
    \begin{itemize}
        \item The answer \answerNA{} means that the paper does not involve crowdsourcing nor research with human subjects.
        \item Including this information in the supplemental material is fine, but if the main contribution of the paper involves human subjects, then as much detail as possible should be included in the main paper. 
        \item According to the NeurIPS Code of Ethics, workers involved in data collection, curation, or other labor should be paid at least the minimum wage in the country of the data collector. 
    \end{itemize}

\item {\bf Institutional review board (IRB) approvals or equivalent for research with human subjects}
    \item[] Question: Does the paper describe potential risks incurred by study participants, whether such risks were disclosed to the subjects, and whether Institutional Review Board (IRB) approvals (or an equivalent approval/review based on the requirements of your country or institution) were obtained?
    \item[] Answer: \answerNA{} 
    \item[] Justification: The paper does not involve human-subject research.
    \item[] Guidelines:
    \begin{itemize}
        \item The answer \answerNA{} means that the paper does not involve crowdsourcing nor research with human subjects.
        \item Depending on the country in which research is conducted, IRB approval (or equivalent) may be required for any human subjects research. If you obtained IRB approval, you should clearly state this in the paper. 
        \item We recognize that the procedures for this may vary significantly between institutions and locations, and we expect authors to adhere to the NeurIPS Code of Ethics and the guidelines for their institution. 
        \item For initial submissions, do not include any information that would break anonymity (if applicable), such as the institution conducting the review.
    \end{itemize}

\item {\bf Declaration of LLM usage}
    \item[] Question: Does the paper describe the usage of LLMs if it is an important, original, or non-standard component of the core methods in this research? Note that if the LLM is used only for writing, editing, or formatting purposes and does \emph{not} impact the core methodology, scientific rigor, or originality of the research, declaration is not required.
    \item[] Answer: \answerYes{} 
    \item[] Justification: LLMs are used as evaluated modules in the agent workflows, as described in the paper and appendix.
    \item[] Guidelines:
    \begin{itemize}
        \item The answer \answerNA{} means that the core method development in this research does not involve LLMs as any important, original, or non-standard components.
        \item Please refer to our LLM policy in the NeurIPS handbook for what should or should not be described.
    \end{itemize}

\end{enumerate}

\end{document}